# Reply to comment on "Perfect drain for the Maxwell fish eye lens".

**Juan C. González, Pablo Benítez, Juan C. Miñano and Dejan Grabovičkić**

Universidad Politécnica de Madrid, Cedint
Campus de Montegancedo
28223 Madrid, Spain

E-mail: jcgonzalez@cedint.upm.es

**Abstract.** We reply to the comments on our paper "Perfect Drain for the Maxwell fish eye lens" (*NJP*. **13** (2011) 023038) made by Fei Sun. We believe that Sun's comments have several mistakes in theoretical concepts and simulation results.

The "Comments on "Perfect Drain for the Maxwel fish eye lens" " made by Fei Sun can be classified in two sections: theoretical and simulation results.

**Theoretical results.**

In relation to the Perfect Drain introduced by Leonhardt in [1] Sun writes: "perfect drain is not only a perfect absorber which can totally absorb all incident radiation without any scattering, but also it can achieve a very sharp electric field around it *which is actually a delta function*". This is not correct: The electric field calculated by Leonhardt in [1] is not an electric field described by delta function. This field is asymptotic in the source and image point, but unlike a delta function, the field is not zero around the points of divergence. An example of the solution described by Leonhardt in [1] is shown in Fig. 1 where the asymptotic behaviour of the function in the neighbourhood of the object and image point can be clearly seen. What is a delta function is the current density at the drain, as it is at the source, and both appear as excitations in the differential equation of the electric field, as was explicitly shown, for instance, in [2]

The system with the drain that we designed in [3] has an electric field distribution identical to that of Leonhardt in Eq. 12 of [1] for all the points outside the drain. When the radius of this drain (which can be arbitrarily selected) tends toward zero, then we get the Leonhardt perfect drain.

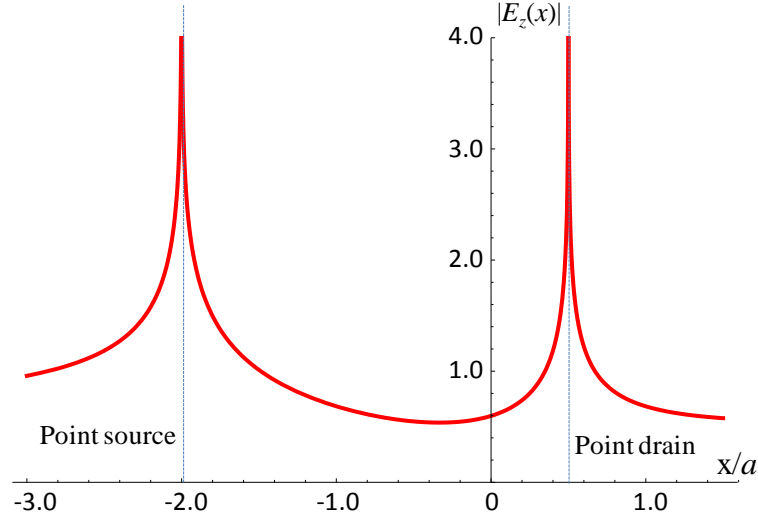

**Fig. 1.** Module of the electric field through the line joining the point source and the point drain. The field diverges in both point sources ($x/a$=-2.0) and drain ($x/a$=0.5).

In the explanation of the results of his simulation Sun writes: "The field distribution in "the perfect drain" can be analytically expressed using Eq. (20) in [18][1]. Even if $\zeta_d$ is very close to 1, $P_{v'}(\zeta)$ does not diverge ($P_{v'}(1)=1$)". Sun is right when he says that $P_{v'}(\zeta)$ does not diverge, but the expression of the electric field inside the drain is given by $E = A\, P_{v'}(\zeta)$. The amplitude $A$ does not depend on the position (i.e, it does not depend on $\zeta$) but is a function of the radius of the drain $\zeta_d$, i.e., $A(\zeta_d)$ and it diverges as the radius of the drain tends toward zero which corresponds to the case $\zeta_d=1$. So the field diverges for the perfect point drain which is the case $\zeta_d=1$. The expression $A(\zeta_d)$ is not explicitly given in our paper [3]. It can be obtained as follows: First calculate $v'$, with the aid of Eq. (22) of our paper [3]. This variable ($v'$) will depend on $\zeta_d$. Then calculate $A$ with any of Eqs. [21] (of the same paper), for instance, with the first one.

$$A(\zeta_d) = \frac{-i e^{i\pi v}\left(P_v(\zeta_d) + i\frac{2}{\pi} Q_v(\zeta_d)\right)}{P_{v'(\zeta_d)}(\zeta_d)} \qquad (1)$$

$v$ is a constant expressed in the Eq. (4) of [3]. The function $P_{v'}(\zeta)$ is not asymptotic but the expression $A(\zeta_d)$ in (1) tends toward infinity when $\zeta_d \to 1$ so the field diverges when the size of the Perfect Drain tends toward zero.

**Simulation.**

The author has carried out the simulation using the following parameters:

$\zeta_d$=0.99, $x_0$=-2 cm, $a$= 1cm, $n_0$=1, and $f$=10GHz.

Where $x_0$ is the position of the source and $a$ and $n_0$ are described by:

$$n(\rho) = \frac{2n_0}{1+(\rho/a)^2} \qquad \rho = \sqrt{x^2 + y^2} \qquad (2)$$

---

[1] Sun's reference [18] is our reference [3].

$n(\rho)$ is the refraction index of the Maxwell fish eye.

We have made the theoretical calculation with these parameters. Fig. 2 shows the module of the electric field as a function of the line joining the source and the image. As can be seen in Fig. 2, the radius of the drain is too high to be considered closed to the Perfect Drain proposed by Leonhardt. Sun should make the simulation with a value $\zeta_d$ closer to 1. Fig. 3 and Fig. 4 show the theoretical electric field distribution for $\zeta_d=0.999$ and $\zeta_d=0.9999$. Clearly, the field has an asymptotic behaviour in the imaging point when $\zeta_d \to 1$.

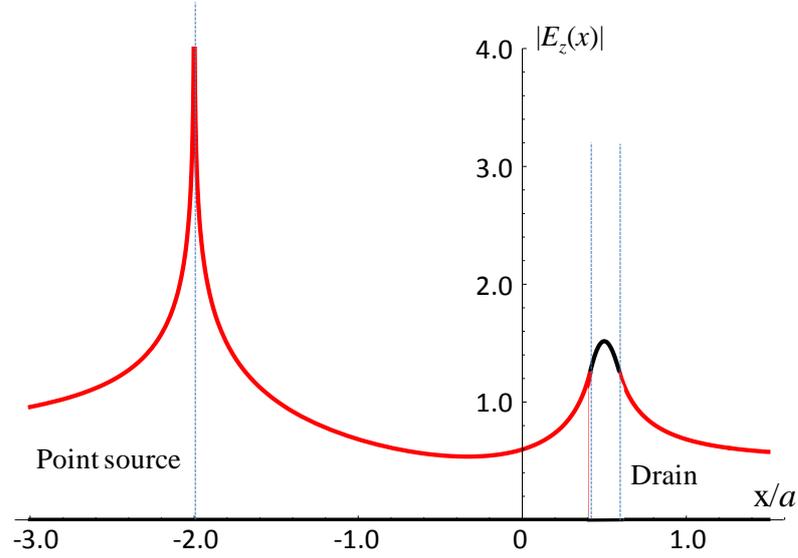

**Fig. 2.** Theoretical module of the electric field for the parameters of the Drain simulated by Fei Sun. The size is far from the punctual Perfect Drain and in consequence the field is not asymptotic in the drain. The parameters obtained with our calculation are: $\zeta_d=0.99$, $\varepsilon_d=8.4544+7.56814i$, RadiusPD/$a$=0.08872.

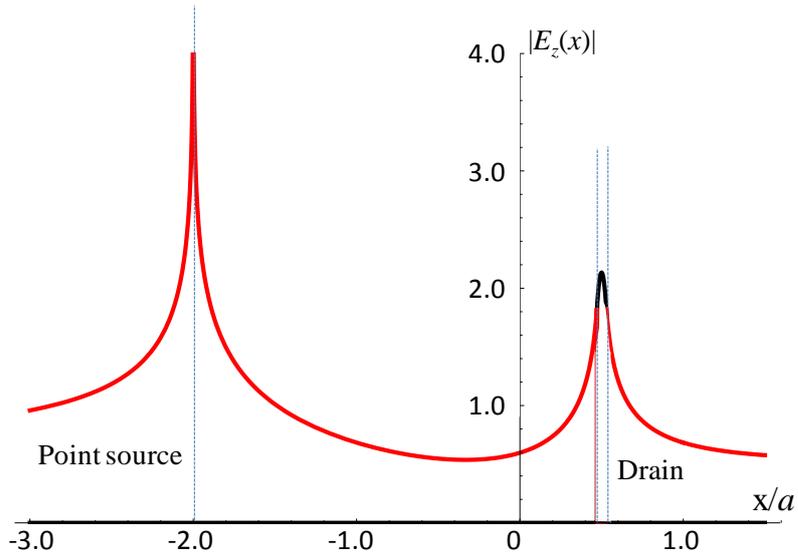

**Fig. 3.** Theoretical module of the electric field for $\zeta_d=0.999$. In this case the radius of the Perfect Drain is smaller and the field distribution more similar to that obtained by Leonhardt shown in Fig. 1. $\varepsilon_d=64.2689+36.693i$, RadiusPD/$a$=0.02796 .

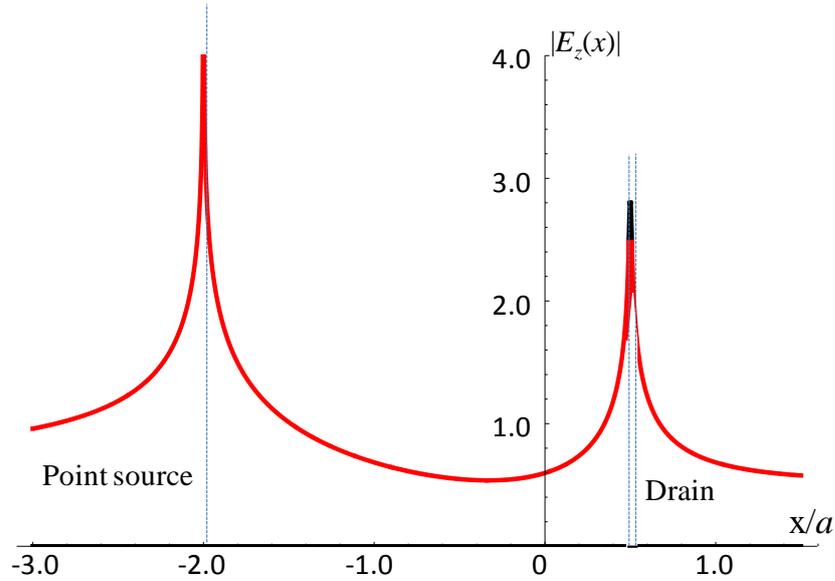

**Fig. 4.** Theoretical module of the electric field for $\zeta_d$=0.9999. The radius of the Perfect Drain is smaller and the field distribution much more similar to the one shown in Fig. 1. $\varepsilon_d$=508.601+206.649i, RadiusPD/$a$=0.008839.

Recent simulation of the Perfect Drain of the Spherical Geodesic Waveguide (SGW) made with COMSOL Multyphisics has been presented in [4] and [5]. The MFE and the SGW are two equivalent systems having the same imaging properties [6]. The results of the simulation show complete agreement with theoretical analysis.

With the parameters used by Sun, we have obtained the dielectric constant, $\varepsilon_d$=8.4544+7.56814i instead of $\varepsilon_d$=1.11089+0.0558956i that Sun reported. Moreover, we have confirmed that the dielectric constant obtained by Sun does not fulfil Eq.22 of [3].

According the graphs presented by Sun, there is no difference in the electric field in two simulations: with and without Perfect Drain, which is clearly a mistake, since there is a reflected wave in both cases (note the difference between the results obtained in Sun's simulation and our theoretical results shown in Fig. 1Fig. 2). The Perfect Drain is a perfect absorber, thus there is no reflection. These results confirm that the $\varepsilon_d$ has not been properly calculated.

We do not understand the values of the x axis in Fig.1. We suppose the variation is between -5.0 and 5.0.

**Conclusion.**

In our opinion this comment needs a rigorous revision, first in relation to the theoretical concepts (the meaning of the perfect drain and the characteristics of the field inside the MFE) and second in relation to the calculation of the parameters for simulation. It seems that the author did not calculate the permittivity of the drain properly. The results of the simulation reflect the mistake.